\documentclass[12pt]{article}
\usepackage{graphicx}
\usepackage{epsfig}

\begin{document}

\title{Relativistic quantum mechanics:\\A Dirac's point-form inspired approach 
} 
\author{ 
B.  Desplanques\thanks{{\it E-mail address:}  desplanq@lpsc.in2p3.fr} \\ 
Laboratoire de Physique Subatomique et de Cosmologie \\
(UMR CNRS/IN2P3--UJF--INPG),\\ 
  F-38026 Grenoble Cedex, France \\  }

\sloppy

\maketitle

\begin{abstract}
\small{
This paper  describes a tentative relativistic quantum mechanics approach 
inspired by Dirac's point-form, which is based on the physics description 
on a hyperboloid surface.  It is mainly characterized by a non-standard 
relation of the constituent momenta of some system to its total momentum. 
Contrary to instant- and front-form approaches, where it takes the form 
of a 3-dimensional $\delta(\cdots)$ function, the relation is given here 
by a Lorentz-scalar constraint. Thus, in the c.m. frame, the sum of the 
constituent momenta,  which differs from zero off-energy shell, has no 
fixed direction, in accordance with the absence of preferred direction 
on a hyperboloid surface. To some extent, this gives rise to an extra 
degree of freedom entering the description of the system of interest. 
The development of a consistent formalism within this picture is described. 
Comparison with other approaches is made. }
\end{abstract} 


\newpage

\section{Introduction}
Among the different forms of relativistic quantum mechanics, the point form 
is the less known one. Following the classification made by Dirac  
\cite{Dirac:1949cp}, this approach implies the description of physics 
on a hyperboloid surface, $x^2=\tilde{\tau}$, which is invariant under 
a Lorentz transformation. This property reflects itself in the construction 
of the Poincar\'e algebra. The four components of the momentum operator, 
$P^{\mu}$, contain the interaction while rotation and boost operators, 
which leave this surface unchanged, are kinematical. 

An implementation of the point-form approach has been considered 
in the literature \cite{Sokolov:1985jv,Lev:1993,Klink:1998}. 
It has been used later on for the 
calculation of form factors in different hadronic systems 
\cite{Allen:1998hb,Allen:2000ge,Wagenbrunn:2000es,Amghar:2003tx,Coester:2003rw} 
as well as theoretical ones \cite{Desplanques:2001zw,Amghar:2002jx}. 
However, as noticed by Sokolov \cite{Sokolov:1985jv}, this ``point-form'' 
approach, where the generators of the Poincar\'e algebra evidence 
the same kinematical or dynamical character as above, differs from Dirac's 
one. It assumes that physics is described on a hyperplane, $v \cdot x=\tau$, 
perpendicular to the velocity of the system \cite{Sokolov:1985jv}. 
The kinematical character of boosts is somewhat trivial in this case 
since, at the same time as the system is boosted, the frame employed 
for the description changes. 

An attempt to justify this approach from considering the description 
of physics on a hyperboloid was recently made \cite{Klink:2000pp},
using approximations however. The contribution of the interaction  
to the momentum operator, $``P^{\mu}\,"$, which is so derived, 
is proportional to the velocity of the system. Actually, this result 
turns out to be an exact one but obtained  from describing physics 
on a hyperplane perpendicular to the velocity of the system. 
One therefore recovers the implementation of the ``point-form'' 
approach that was recognized by Sokolov as being different from 
Dirac's point-form  \cite{Sokolov:1985jv}. An implementation 
of this last approach therefore remains an open problem.

In the present work, we propose ourselves to make some steps in this 
direction. Each approach is, in particular, characterized by the relation 
of the constituent momenta to the total momentum of the system under 
consideration. In the instant- or front-form 
approaches for instance, the sum of the constituent momenta 
is equal to the total momentum carried by the system (instant-form) 
or deviates from it by an amount proportional to the orientation 
of the hyperplane which the physics is described on (front-form). 
However, for the Dirac's point form, where no orientation in Minkowski 
space is a priori preferred, the relation must necessarily take 
the form of a Lorentz-scalar constraint. The simplest one that can 
thus be imagined for a two-body system, compatible with the fact 
that the standard 3-momentum conservation should be recovered in the  
non-relativistic limit (small binding), takes the form  
\begin{equation} 
(\vec{p}_1+\vec{p}_2-\vec{P})^2-(e_1+e_2-\;E_P)^2=-(p_1+p_2-P)^2=0 \, .
\label{int1}
\end{equation}

Such a constraint implies that the sum of the constituent momenta 
deviates from the total momentum of the system like in other approaches 
(front form) but, most importantly, the departure  takes the form 
of a vector whose orientation is not determined. This introduces a 
somewhat new, non-trivial, degree of freedom in describing a given system. 
Whether this can be implemented consistently in some 
formalism is not straightforward. Difficulties arise as soon as one 
considers the equation that should be fulfilled by the wave function 
describing the system. Everything concerned with the total momentum, 
$\vec{P}$, or the sum of the constituent momenta, $\vec{p}_1+\vec{p}_2$, 
should factor out so that the solution can be expressed in terms of 
the solution of a mass equation involving only internal variables. 
Thus, it has to be shown that the constraint given by Eq. (\ref{int1}) 
is part of the wave function. We mentioned that the direction of 
the momenta carried by the constituents is not fixed. How this direction, 
which enters the wave function, is affected by the interaction is another 
important aspect. Lastly, there is the possibility that the  interaction 
itself depends on the total momentum but this problem is no more than 
the one appearing in instant- and front-form approaches. How to solve 
this problem was considered by Bakamjian and Thomas \cite{Bakamjian:1953kh} 
(see also Ref. \cite{Amghar:2002jx} for a practical application 
of interest here). 

The plan of the paper is as follows. In the second section, we describe 
how we can disentangle the external and internal degrees of freedom, 
with the aim to derive a mass operator involving only the last ones. 
This is achieved by making an appropriate change of variables and 
showing that the orientation taken by the sum of the constituent 
momenta with respect to the total momentum, $\vec{P}$, despite 
it is not determined, is nevertheless conserved by the interaction. 
The third section is devoted to a comparison with other forms. 
In particular, we briefly show how the present implementation of 
the point-form approach corrects for the systematic failure of an 
earlier version \cite{Klink:1998} to provide the expected asymptotic 
power law behavior of form factors. A conclusion and further 
discussion is given in the fourth section. As the approach presented 
here rather involves new aspects, many details are important. 
These ones are given in the appendix, keeping only in the main 
text the minimal ingredients required for its understanding.

\section{Equation for the wave function and mass operator}
In this section, we determine the expression that the wave function 
of a two-body system composed of scalar constituents could take 
in an approach inspired from Dirac's point-form. The relation 
to a mass operator that could be used in other forms, or taken 
from them, is made. 

\begin{figure}[htb]
\begin{center}
\mbox{ \psfig{ file=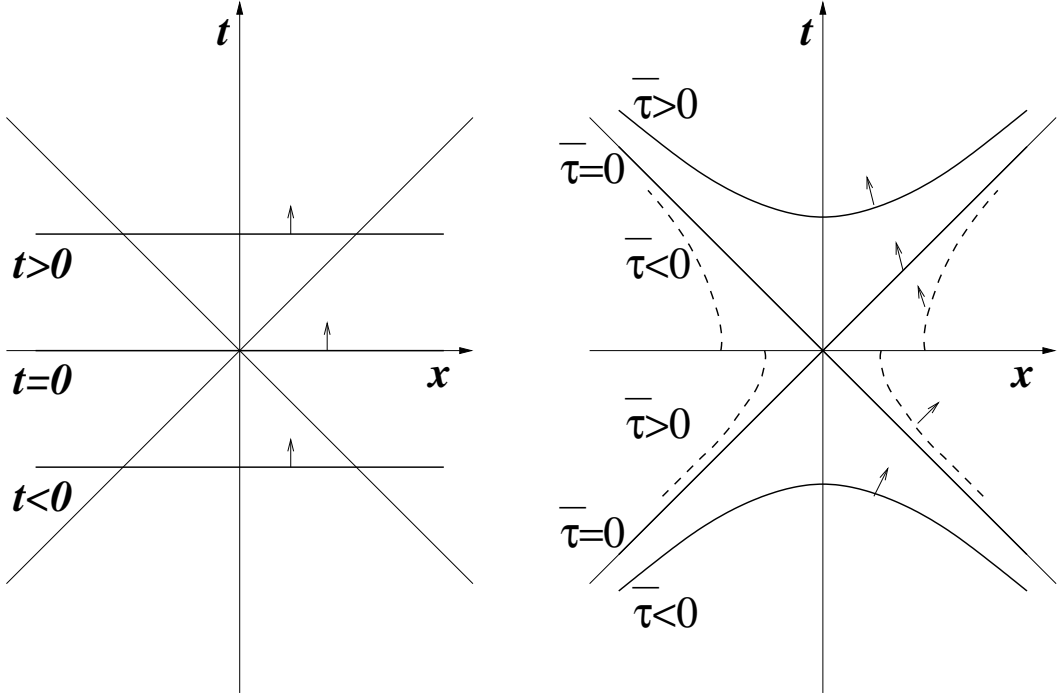, width=14cm}}
\end{center}
\caption{Transverse representation of hypersurfaces at different ``times'' 
in both instant-and point-form approaches (left and right parts respectively). 
The last one is given for the choice, $U^{\mu}=1, \,0 ,\,0, \,0$. For
convenience, the variable $\tilde{\tau}$ is denoted $\bar{\tau}$. 
The arrows indicate how the hypersurfaces evoluate when the ``time'' 
varies from $-\infty$ to $+\infty$. The counterpart of the hyperboloid 
surface with a space-like character, which is rarely mentioned and 
corresponds to a different sign of the ``time'' $\tilde{\tau}$, 
is represented by the dashed curve.
\label{fig0} }
\end{figure}

\noindent
$\bullet$ {\bf Relation to a description on a hyperboloid surface}\\
The derivation of an interaction operator may start from the current 
expression of the 4-momentum, $P^{\mu}$, in terms of the free-particle 
momenta and the Lagrangian interaction density:
\begin{equation} 
``P^{\mu}\,"= ``p^{\mu}\,"+``P^{\mu}_{int}"=
``p^{\mu}\,"- \int d^4x \;g^{\mu \nu} f_{\nu}(x) \; {\cal L}_{int} (x) \, ,
\label{wf1}
\end{equation}
where quotation marks have been introduced to distinguish the operator 
and the corresponding eigenvalue. Extra terms involving the field derivatives 
don't need to be considered here. 
The function $f_{\nu}(x)$ involves the hypersurface which the integral 
is performed over. It takes different expressions depending on the form 
under consideration. In the case of a hyperplane defined by the orientation
$\lambda_{\nu}$, the general function is given by 
$f^{\tau}_{\nu}(x)=\lambda_{\nu}\; \delta(\lambda \! \cdot \! x-\tau)$.  
In the present case, the expression of interest, whose derivation is given 
in appendix \ref{app:a}, reads:
\begin{equation} 
 f^{\tilde{\tau}}_{\nu}(x)= 2\, x_{\nu}\; 
\Big( \theta( U \!\cdot\! x)\;
\delta(\tilde{\tau}-x^2 )
- \theta( -U \!\cdot\! x)\;
\delta(\tilde{\tau}+x^2 ) \Big)+\cdots\, ,
\label{wf2}
\end{equation}
where the $\delta(\tilde{\tau}-x^2)$ function, for instance, implies 
that the hypersurface to be integrated over is a hyperboloid 
(for $\tilde{\tau}>0$). 
The $\theta(\cdots)$  functions 
indicate how the two parts of the hypersurface contribute, the relative 
sign, in relation with space-time symmetry, being essential to get 
a meaningful contribution (see appendix \ref{app:a}). They involve 
a 4-vector, $U^{\mu}$, that will be defined later on in terms 
of the momenta of particles participating to the interaction. 
This choice is the simplest one which is possible within the point-form 
approach so that to ensure the Lorentz invariance of the argument 
of the $\theta(\cdots)$ functions. The above expression of 
$f^{\tilde{\tau}}_{\nu}(x)$ also evidences a dependence on the 
$\tilde{\tau}$  variable, which plays a role analogous to the time 
in the instant form. It can be used to specify the hypersurface of interest 
and is such that the whole space-time volume is covered when it varies 
between its extreme limits ($-\infty$ and $+\infty$ here). Any value of 
this ``time" should be acceptable but corresponding mathematical developments 
are generally quite involved. Due to obvious ``time"-symmetry reasons, 
some simplification is however expected with the particular choice 
considered by Dirac \cite{Dirac:1949cp}, $\tilde{\tau}=0$, 
where the hyperboloid surface 
reduces to the light cone while the contribution represented by dots 
in Eq. (\ref{wf2}) cancels. This is similar to making the current choice 
$t=0$ in the instant form, where the time evolution could be accounted 
for by an appropriate equation involving the Hamiltonian. In order to provide 
a comparison, the two cases are illustrated in Fig. \ref{fig0} for various
values of the ``time'' variable. Paying attention 
to how the limit $\tilde{\tau} \rightarrow 0$ should be taken (see appendix 
\ref{app:a}), the relevant function $f_{\nu}(x)$ that we will use reads:
\begin{equation} 
f_{\nu}(x)= x_{\nu} \; \delta(x^2)\; 
\Big[\theta(U \! \cdot  \!x)-  \theta(-U  \!\cdot \! x)\Big]\, .
\label{wf3}
\end{equation}

In field theory, the Lagrangian density ${\cal L} (x)$ would have a 
well-defined meaning. In relativistic quantum mechanics, this density has 
to be re-interpreted for a part. On the one hand, the $x$ coordinate refers 
to the system as if this one could be considered as an elementary one. 
On the other hand, the density accounts for degrees of freedom that are 
integrated out, like meson exchanges, leading to a mass operator 
describing the internal structure of the system of interest. It also 
involves the plane waves describing the interacting particles. 
These various inputs can be considered as relevant to a minimal 
space-time description of relativistic quantum mechanics. They allow 
one to recover the usual expressions pertinent to the implemention 
of instant- and front-forms approaches. As an example of interest 
in the present work, we write the interaction contribution 
to $``P^{\mu}\,"$, Eq. (\ref{wf1}), for spinless-scalar particles 
exchanging a spinless meson of mass $\mu$:
\begin{eqnarray}
``P^{\mu}_{int}"=- 
\int  d\vec{p}_1 \; d\vec{p}_2 \; d\vec{p}\,'_1 d\vec{p}\,'_2
\int d^4x \; x^{\mu} \; \delta(x^2)\; 
\Big[ \theta(U \! \cdot  \!x)-  \theta(-U \! \cdot \!  x) \Big]
\; e^{i(p-p')  \cdot  x}
\nonumber \\ \times
\frac{ a^*(\vec{p}_1) \; a^*(\vec{p}_2  )}{ 
(2\,\pi)^3\; (2\,e_1)^{1/2}\;(2\,e_2)^{1/2}}\;\;
\frac{4\,m^2\;g^2}{\mu^2+\cdots} \;\;
\frac{a(\vec{p}\,'_1) \; a(\vec{p}\,'_2) 
}{(2\,\pi)^3\;(2\,e'_1)^{1/2}\;(2\,e'_2)^{1/2}}\, .
\label{wf4}
\end{eqnarray}
While standard notations are mostly employed, we stress that 
the quantity $p^{\mu}$ is a shortened notation to represent  
the sum of the momenta relative 
to particles 1 and 2, $p^{\mu}= p^{\mu}_1+p^{\mu}_2$.  We notice that 
the integral over $x$ replaces the one which provides the 3-momentum 
conservation in the instant form. Also, the momentum transfer carried 
by the exchanged meson, that appears at the denominator of its 
propagator, has been purposely replaced by dots. As will be seen below, 
this term could  be affected by consistency conditions pertinent to a 
relativistic quantum mechanics approach. For an illustration purpose, the 
interaction has been given the form of a single-boson exchange but 
it could already have at this point an effective character and account 
for some field-theory corrections through the choice of the coupling 
or that of the boson mass \cite{Amghar:2000pc}.

\noindent
$\bullet$ {\bf Basic equation} \\
Instead of looking for a solution of Eq. (\ref{wf1}), we consider 
an equivalent problem that consists in searching for a solution of the 
following equation:
\begin{equation}
P^2\;\Phi_P(\vec{p}_1,\vec{p}_2)=
\Big(p^2+p \! \cdot \! P_{int}+ P_{int} \! \cdot \!  p
+P_{int}^2 \Big)\;\Phi_P(\vec{p}_1,\vec{p}_2) \, ,
\label{wf5}
\end{equation}
where $\Phi_P(\vec{p}_1,\vec{p}_2)$ is the two-body wave function. 
This equation offers the advantage to be closer to a Lorentz-invariant 
mass equation, facilitating the derivation of a mass operator. 
In the following, we ignore the last term at the r.h.s. 
($P_{int}^2 $). Quite generally, its contribution could be incorporated 
in the middle terms (from a phenomenological view point). 
In the present case however, it will turn out to vanish, very much 
like in front-form approaches where $P^{\mu}_{int}$ is proportional 
to a 4-vector $\omega^{\mu}$ with the property $\omega^2=0$ 
(see for instance Ref. \cite{Carbonell:1998rj}). 

After replacing $P^2$ in Eq. (\ref{wf5}) by the eigenvalue $M^2$, 
and using the expression of the integral over $x$ appearing 
in Eq. (\ref{wf4}), given by Eq. (\ref{appa12}) of appendix \ref{app:a}, 
we can now cast the above equation into the following basic form:
\begin{eqnarray}
\nonumber 
&&\hspace{-0.5cm}\Big(M^2-p^2\Big)\;\Phi_P(\vec{p}_1,\vec{p}_2)
\nonumber \\ && 
= - 
\int  \int 
\frac{d\vec{p}\,'_1}{(2\,\pi)^3} \; \frac{d\vec{p}\,'_2}{(2\,\pi)^3} \;\;    
\frac{1}{(2\,e_1)^{1/2} \;(2\,e_2)^{1/2}}\;\;
\frac{4\,m^2\;g^2}{\mu^2+\cdots} \;\;
\frac{1}{(2\,e'_1)^{1/2} \;(2\,e'_2)^{1/2}}\;
\nonumber \\ && \hspace{0.4cm} \times \, 
(p+p')\!\cdot\! \; \partial_{p-p'} \Bigg(
4\,\pi^2\; \delta\Big((p-p')^2\Big)\; \Big[ 
\theta\Big(U \!\cdot\! (p-p') \Big)- \theta\Big(U \!\cdot\! (p'-p) \Big)
\Big] \Bigg)\; \Phi_P(\vec{p}\,'_1,\vec{p}\,'_2) \, . 
\nonumber \\ 
\label{wf6} 
\end{eqnarray} 

An equation whose Lorentz invariance is more transparent for some factors is 
obtained by making the change:
\begin{equation}
\Phi_P(\vec{p}_1,\vec{p}_2)=
\frac{1 }{ (2\,e_1)^{1/2} \;(2\,e_2)^{1/2} } 
\;\Phi^{(1)}_P(\vec{p}_1,\vec{p}_2) \, .
\label{wf7}
\end{equation} 
Together with making the derivative in the integrand, the equation reads: 
\begin{eqnarray}
&& \hspace{-0.5cm}\Big(M^2-p^2\Big)\;\Phi^{(1)}_P(\vec{p}_1,\vec{p}_2)
= - \frac{1}{8\,\pi^4}
\int  \int 
\frac{d\vec{p}\,'_1}{2\,e'_1} \; \frac{d\vec{p}\,'_2}{2\,e'_2} \;\;    
\frac{4\,m^2\;g^2}{\mu^2+\cdots} \;
\nonumber \\ &&  \hspace{0.1cm} \times
\Bigg(U \!\cdot\! (p+p')\;\delta\Big((p-p')^2\Big)\;
\delta\Big(U \!\cdot\! (p-p')\Big) 
\nonumber \\ && \hspace{0.7cm}
+\,(p^2-p'^2) \; \delta'\Big((p-p')^2\Big)\; \Big[ 
\theta\Big(U \!\cdot\! (p-p') \Big)- \theta\Big(U \!\cdot\! (p'-p) \Big)
\Big] \Bigg)\; \Phi^{(1)}_P(\vec{p}\,'_1,\vec{p}\,'_2) \, .
\label{wf8} 
\end{eqnarray} 
We now want to see whether, or under which conditions, the above equation
admits a mass spectrum independent of the total momentum, $\vec{P}$. 
It is stressed that it does not involve at the r.h.s. 
any 3-dimensional $\delta(\cdots)$ function ensuring the conservation 
of some momentum like in the instant- or front-form formalisms. 
For the sake of comparison, we here give this equation for the 
general case of a hyperplane of orientation $\lambda^{\mu}$:
\begin{eqnarray}
&& \hspace{-0.5cm}\Big(M^2-p^2\Big)\;\Phi^{(1)}_P(\vec{p}_1,\vec{p}_2)
= - \frac{1}{(2\,\pi)^3}
\int  \int 
\frac{d\vec{p}\,'_1}{2\,e'_1} \; \frac{d\vec{p}\,'_2}{2\,e'_2} \;\;    
\frac{4\,m^2\;g^2}{\mu^2+\cdots} \;
\nonumber \\ &&  \hspace{2cm} \times
\frac{\lambda \!\cdot\! (p+p')}{\lambda^0}\; 
\delta\Big(\vec{p}-\vec{p}\,'
-\frac{\vec{\lambda}}{\lambda^0}\;(e_p-e_{p'})\Big)\;
 \Phi^{(1)}_P(\vec{p}\,'_1,\vec{p}\,'_2) \, .
\label{wf8bis} 
\end{eqnarray} 
It is noticed that, in the limit $|\vec{\lambda}/ \lambda^0| \rightarrow 1$ 
(front-form case), the 3-dimensional $\delta(\cdots)$ function
in this last equation implies those present in the previous one  
(assuming also $\lambda^{\mu} \propto U^{\mu}$). The converse result 
does not hold however. 

\noindent
$\bullet$ {\bf Essential aspects of a solution} \\
To obtain a solution of the above equation, we first assume that the 
4-vector, $U^{\mu}$, can be written as:
\begin{equation}
U^{\mu}=c\;(p-P)^{\mu}+ c'\;(p'-P)^{\mu} \, ,
\label{wf9}
\end{equation}
where  $c$ and $c'$ are arbitrary coefficients at this point. It is then 
possible to show the following relation (see appendix \ref{app:b}): 
\begin{eqnarray}
&&\hspace{-0.5cm}\Bigg(U \!\cdot\! (p+p')\;\delta\Big((p-p')^2\Big)\;
\delta\Big(U \!\cdot\! (p-p')\Big) 
\nonumber \\ && \hspace{-0.1cm}
+\,(p^2-p'^2) \; \delta'\Big((p-p')^2\Big)\; \Big[ 
\theta\Big(U \!\cdot\! (p-p') \Big)- \theta\Big(U \!\cdot\! (p'-p) \Big)
\Big] \Bigg) \; \delta\Big((p'-P)^2\Big)
\nonumber \\ && \hspace{-0.5cm}=
\delta\Big((p-P)^2\Big) \; \delta\Big((p-P) \!\cdot\! (p'-P)\Big) \; 
\delta\Big((p'-P)^2\Big)\;
\frac{c\;(p'^2-M^2)+c'\;(p^2-M^2)}{c+c'} \, .
\label{wf11} 
\end{eqnarray} 
An important consequence of this result is that, if the wave 
function appearing on the r.h.s. of Eq. (\ref{wf8}) is proportional to 
$\delta\Big((p'-P)^2\Big)$, then, the wave function appearing 
on the l.h.s. is proportional to $\delta\Big((p-P)^2\Big)$. 
This factor is therefore part of the solution that is looked for. 
Introducing the change:
\begin{equation}
\Phi^{(1)}_P(\vec{p}_1,\vec{p}_2)=
\delta\Big((p-P)^2\Big)
\;\Phi^{(2)}_P(\vec{p}_1,\vec{p}_2) \, ,
\label{wf12}
\end{equation} 
a further reduction of Eq. (\ref{wf8}) is possible:
\begin{eqnarray}
&&\Big(M^2-p^2\Big)\;\Phi^{(2)}_P(\vec{p}_1,\vec{p}_2)
= - \frac{1}{8\,\pi^4}
\int  \int 
\frac{d\vec{p}\,'_1}{2\,e'_1} \; \frac{d\vec{p}\,'_2}{2\,e'_2} \;\;    
\frac{4\,m^2\;g^2}{\mu^2+\cdots} \;
\nonumber \\ &&  \hspace{0.5cm} \times \,
\delta\Big((p-P) \!\cdot\! (p'-P)\Big) \; 
\delta\Big((p'-P)^2\Big)\;
\frac{c\;(p'^2-M^2)+c'\;(p^2-M^2)}{c+c'} \; 
\Phi^{(2)}_P(\vec{p}\,'_1,\vec{p}\,'_2) \, . 
\nonumber \\
\label{wf13} 
\end{eqnarray} 
From examining this last equation, another feature comes out. It concerns 
the orientation  of the 3-vectors, $\vec{p}-\vec{P}$ and $\vec{p}\,'-\vec{P}$. 
Taking into account that $(p-P)^2=(p'-P)^2=0$, it is appropriate 
to introduce unit vectors defined as:
\begin{equation}
\vec{u}= \frac{\vec{p}-\vec{P}}{e-E_P}\, , \;\;\; 
\vec{u}\,'= \frac{\vec{p}\,'-\vec{P}}{e'-E_P}\, .
\label{wf14}
\end{equation} 
These vectors determine for a part how the momentum carried by the 
constituents depart from the total momentum carried by the system. 
The presence of the function $\delta\Big((p-P) \cdot (p'-P)\Big)$ 
in Eq. (\ref{wf13}) then implies that the angle between the two 3-vectors, 
$\vec{u}$ and $\vec{u}\,'$ is zero (see appendix \ref{app:b}). Therefore, 
the orientation of this vector is not affected by the interaction. 
It remains conserved through the rescattering processes that solving 
Eq. (\ref{wf13}) implies, with the result: 
\begin{equation}
\vec{u}= \frac{\vec{p}-\vec{P}}{e-E_P}= 
\frac{\vec{p}\,'-\vec{P}}{e'-E_P}= 
\frac{\vec{p}\,''-\vec{P}}{e''-E_P}= \cdots \, .
\label{wf15}
\end{equation} 

At this point, the nature of the 4-vector $U^{\mu}$ introduced 
at the beginning of this subsection can be made more precise. 
As can be checked from Eq. (\ref{wf6}) for instance, present results 
do not depend on its scale. Moreover, as a consequence of the three 
$\delta(\cdots)$ functions appearing on the r.h.s. of Eq. (\ref{wf11}), 
it has  a zero norm. The 4-vectors $U^{\mu}$ and $(1,\; \vec{u})$ 
can therefore be  identified up to a factor. It is also noticed that 
the 4-vector $U^{\mu}$ offers many similarities with the 4-vector, 
$\omega^{\mu}$, which determines the orientation of the front 
in front-form approaches. In this case, an equation like Eq. (\ref{wf15}) 
is obtained but with $\vec{u}$ replaced by the orientation of the front, 
$\vec{n}$  (see for instance Ref. \cite{Carbonell:1998rj}). 
These similarities may be useful to understand the relationship between 
different approaches. However, there are major differences. 
The orientation $\vec{u}$ has to be integrated over while $\vec{n}$ 
is a fixed orientation. On the other hand, the unit character of $\vec{u}$ 
is obtained from a consistency requirement while, for $\vec{n}$, 
it is given as part of the formalism. Finally, but not independently, 
the underlying formalisms imply physics description on quite different 
hypersurfaces (light cone in one case and hyperplane tangent to it 
in the other).

\noindent
$\bullet$ {\bf Change of variables} \\
The next step in dealing with Eq. (\ref{wf6}) is to show that its mass 
spectrum is independent of the total momentum, $\vec{P}$, or to 
determine under which conditions this is realized. This can be 
achieved by reducing this equation to a unique one, independent 
of the total momentum. In this order, we follow the approach 
pioneered by Bakamjian and Thomas \cite{Bakamjian:1953kh}, with 
appropriately adapting their work for the instant-form case 
to the present one. 

We first make a change of variables 
which, for one-particle states, preserves the relation $p^2_{1,2}=m^2$. 
Not surprisingly, it is taken here as a
Lorentz-type transformation. For a part, it contains the standard 
Lorentz transformation which relates a system with a finite momentum 
to that one with zero momentum. For another part, it contains 
a Lorentz-type transformation  which relates momenta of constituents 
in the center of mass to internal variables, $\vec{k}$ and $\vec{u}$. 
Moreover, we require that Eq. (\ref{wf15}) be automatically satisfied. 
For particle 1, the change of variable thus takes the form: 
\begin{eqnarray}
\vec{p}_1 &=& \vec{k} +\vec{w}\;\frac{\vec{w} \! \cdot \!
\vec{k} }{\sqrt{1+\vec{w}^2} +1}
 + \vec{w}  \;e_k \, ,
\nonumber \\
e_1 &=& \sqrt{1+\vec{w}^2}\;e_k+ \vec{w}\! \cdot \!
 \vec{k}   \, .
\label{wf16}
\end{eqnarray}
Expressions for particle 2 are obtained by making in these equations 
the change: $\vec{k} \rightarrow -\vec{k}$. 
From them, one  easily obtains the following equalities: 
\begin{eqnarray}
\vec{p}=\vec{p}_1+ \vec{p}_2&= &2\,e_k\;\vec{w} \, ,
\nonumber \\
e=e_1+e_2&=&2\,e_k\;\sqrt{1+\vec{w}^2}\, .
\label{wf17}
\end{eqnarray}
Equation (\ref{wf15}) is then identically fulfilled for any $\vec{u}$  
(including $|\vec{u}| \neq 1$) by taking:  
\begin{eqnarray}
\vec{w} &=& \frac{ \vec{P} }{ 2\,e_k }
+ \frac{ \vec{u} }{ 2\,e_k } \; \frac{ 4\,e_k^2-M^2 }{u \! \cdot \!P
+\sqrt{ (u \! \cdot \!P)^2 +(1-\vec{u}\,^2)\,(4\,e_k^2-M^2 )  } }, 
\nonumber \\
\sqrt{1+\vec{w}^2} &= & \frac{E_P }{ 2\,e_k } 
+\frac{1}{2\,e_k} \; \frac{4\,e_k^2-M^2 }{u \! \cdot \!P
+\sqrt{ (u \! \cdot \!P)^2 +(1-\vec{u}\,^2)\,(4\,e_k^2-M^2 )  }  } \; .
\label{wf18}
\end{eqnarray}

In a next step, the above changes are made in Eq. (\ref{wf13}), which 
then involves an integral over $\vec{k}\,'$ and $\vec{w}\,'$. 
Taking advantage of the two $\delta(\cdots)$ functions that appear in 
this equation, the second of the integral can be easily performed (see 
some details in appendix \ref{app:c}).  The following equation for 
$\Phi^{(2)}_P(\vec{p}_1,\vec{p}_2) $ is thus obtained: 
\begin{eqnarray}
\Big(M^2-4\;e_k^2\Big)\; \Phi^{(2)}_P(\vec{p}_1,\vec{p}_2)
= \int  
\frac{d\vec{k}\,'}{(2\pi)^3\;e_{k'}}\; 
\Big( \frac{-4\,m^2\;g^2}{\mu^2+\cdots} \Big)\;
\frac{\sqrt{4\,e_{k'}^2-M^2} }{ \sqrt{4\,e_k^2-M^2} }\; \; \hspace{2cm}
\nonumber \\        \times
\Bigg( \frac{ c\;(4\,e_{k'}^2-M^2)+c'\;(4\,e_k^2-M^2) }{ \Big( c+c' \Big) 
 \sqrt{ 4\,e_k^2-M^2}\;\sqrt{4\,e_{k'}^2-M^2} } \Bigg)\;
 \Phi^{(2)}_P(\vec{p}\,'_1,\vec{p}\,'_2)\, .
\label{wf19}
\end{eqnarray} 

Assuming for a while that one can forget about the meson propagator, 
this equation would have the desired properties. The solutions 
do not depend on the total momentum, $\vec{P}$, and only involve 
the internal variable, $\vec{k}$. Referring to the solution of an equation 
that is more symmetrical in variables  $\vec{k}$ and $\vec{k}\,'$: 
\begin{eqnarray}
&&(M^2-4\,e_k^2)\;\phi_0(\vec{k})=
\int \frac{d\vec{k}\,'}{(2\pi)^3} \; 
\frac{1}{ \sqrt{e_{k}} }\; \frac{-4\,m^2\;g^2}{\mu^2+\cdots}\;
  \frac{1}{ \sqrt{e_{k'}} } \; \; \phi_0(\vec{k}\,')
\nonumber  \\ && \hspace{4cm}\times
\Bigg (\frac{c\;(4\,e_{k'}^2-M^2)+c'\;(4\,e_k^2-M^2) }{
 (c +c')\;\sqrt{4\,e_k^2-M^2}\;\sqrt{4\,e_{k'}^2-M^2} }
\Bigg) \, ,
\label{wf21}
\end{eqnarray} 
the solution of Eq. (\ref{wf19}) may read: 
\begin{equation}
\Phi^{(2)}_P(\vec{p}_1,\vec{p}_2) =
\frac{\sqrt{e_k}\;\phi_0(\vec{k})}{\sqrt{4\,e_k^2-M^2}} \, .
\label{wf22}
\end{equation} 
Thus, apart from the interaction term represented by a meson propagator 
in  Eqs. (\ref{wf6}), (\ref{wf8}), (\ref{wf13}), or (\ref{wf19}), 
which was not considered explicitly till here, it has been possible 
to get rid of their dependence on the total momentum, $\vec{P}$.

When the change of variable made above is applied to the meson 
propagator, it is found that the term corresponding to the usual 
squared momentum transfer, represented by dots in the above equations, 
depends on the $\vec{k}$ variable, but also on the total momentum, 
possibly through the $\vec{u}$ variable. 
Removing this total momentum dependence represents the third step 
in getting a relevant mass operator. The problem was considered in 
Ref. \cite{Amghar:2002jx} with some details. We therefore only remind 
the main point underlying the solution. This total momentum-dependent 
term also evidences an off-shell character and, thus, its contribution 
has the same order as higher-order terms in the interaction. It is 
expected that adding these terms to the interaction kernel 
should restore the total momentum independence, with the consequence 
that the whole interaction, necessarily effective, only depends 
on the $\vec{k}$ variable. We thus recover the same kind of constraint 
as the one emphasized first in the instant-form case by Bakamjian 
and Thomas \cite{Bakamjian:1953kh} and in other forms since then 
(see Ref. \cite{Keister:sb} for a review). The expression of the 
mass operator for the present two-body system may then read:
\begin{equation}
M^2=4\,e_k^2+4\,m\;\tilde{V} \, ,
\label{wf23}
\end{equation}
where $\tilde{V}$ only depends on the internal variable $\vec{k}$ 
and is normalized like the interaction appearing in a standard 
Schr\"odinger equation. For the one-boson exchange contribution 
considered above, this interaction in momentum space up to off-shell 
effects thus reads:
\begin{equation}
\tilde{V}(\vec{k},\vec{k}\,')=  -
\sqrt{\frac{m}{e_k}} \;\; \frac{g^2}{\mu^2+(\vec{k}-\vec{k}\,')^2} \;\; 
\sqrt{\frac{m}{e_{k'}}} \, .
\label
{wf24}
\end{equation}
Of course, the interaction can contain higher order terms or, if a 
phenomenological approach is used, the parameters, such as the coupling 
constant $g^2$, could be fitted to some data, accounting for higher-order 
effects as already mentioned. 

\noindent
$\bullet$ {\bf Off-shellness of the interaction}\\
Examination of Eq. (\ref{wf21}) shows the presence of a factor 
that evidences a dependence on the definition of the 4-vector $U^{\mu}$ 
through the dependence on the coefficients, $c$ and $c'$. It is noticed 
that this factor is equal to 1 on energy shell, independently of the values 
of $c$ and $c'$. It therefore involves off-shell effects. As such, 
its effect cannot be distinguished from the usual off-shellness 
ambiguitites of the interaction. It could therefore be absorbed into the 
determination of the effective interaction. The off-shellness behavior 
of the factor is not quite usual however. Whether this feature has a deep 
meaning is unclear. On the other hand, it is always possible to choose 
the coefficients $c$ and $c'$ in such a way that the factor be equal 
to 1, thus eliminating the problem if any. The corresponding 
expressions are determined up to a common factor:
\begin{equation} 
c \propto (4\,e_k^2-M^2)^{1/2},\;\;\; c' \propto (4\,e_{k'}^2-M^2)^{1/2} \, .
\label{wf25}
\end{equation}
%

\noindent
$\bullet$ {\bf Normalization and orthogonality of the solutions}\\
The normalization of the solutions of the mass operator, Eq. (\ref{wf23}), 
labelled by an index $\alpha$, may be given by:
\begin{equation}
\int \frac{d\vec{k}}{(2\pi)^3} \; 
\phi^{\alpha}_0(k) \; \phi^{\alpha'}_0(k)
=N^{\alpha}\;\delta^{\alpha,\alpha'}  \, .
\label{wf26}
\end{equation}
The derivation of the above expression is a standard one. It consists in  
sandwiching the mass operator, Eq. (\ref{wf23}), between two solutions, 
possibly different, and making the operator to act respectively on the 
right and on the left. By taking the difference of the results so obtained, 
one gets zero at the r.h.s. using the hermiticity of the operator,  
while the l.h.s. is given by: 
\begin{equation}
\Big(M_{\alpha}^2-M_{\alpha'}^2\Big)
\int \frac{d\vec{k}}{(2\pi)^3} \;\phi^{\alpha}_0(k) \; \phi^{\alpha'}_0(k)
=0  \, .
\label{wf27}
\end{equation}
This equation implies that the integral is zero if the masses are different, 
or a constant otherwise, as given by Eq. (\ref{wf26}). For our purpose, 
this result has to be extended to take into account that solutions we are
interested in also depend on the total momentum of the system and the 
4-vector $U^{\mu}$ or $(1,\; \vec{u})$. Applying the same method for Eq. 
(\ref{wf8}) and noticing that the part involving the internal variable 
$\vec{k}$ factors out, one gets the following expression for the 
normalization: 
\begin{eqnarray}
N(\vec{P}_{\alpha},\vec{P}_{\alpha'}) & = & \frac{1}{(2\pi)^6}
\int \frac{d\vec{p}_1}{2\,e_1} \;\; 
\frac{d\vec{p}_2}{2\,e_2}\;
\Phi^{(2)}_{P_{\alpha}}(\vec{p}_1,\vec{p}_2)\;M_{\alpha}\;
\sqrt{p^2-M_{\alpha}^2}
\nonumber \\
&& \times  \; 8\,\pi \;\delta\Big((p-P_{\alpha})^2\Big)
\; \delta\Big((p-P_{\alpha}) \!\cdot\!(p-P_{\alpha'})\Big)\; 
\delta\Big((p-P_{\alpha'})^2\Big)\;
\nonumber \\
&& \times \; \Phi^{(2)}_{P_{\alpha'}}(\vec{p}_1,\vec{p}_2)\;
M_{\alpha'}\;\sqrt{p^2-M_{\alpha'}^2}
\nonumber \\
& = &  E_{P_{\alpha}}\; 
\delta\Big(\vec{P}_{\alpha}-\vec{P}_{\alpha'}\Big)\;\;
  N^{\alpha}\;\delta^{\alpha,\alpha'} \;
 \int \frac{d\vec{u}}{2\,\pi } \;\delta(1-\vec{u}\,^2)\;
  \frac{M_{\alpha}^2}{(u \!\cdot\! P_{\alpha})^2}\, .
\label{wf28}
\end{eqnarray}
Some details about dealing with the $\delta(\cdots)$ functions appearing 
in the above equation are given in appendix \ref{app:d}.
We notice that the last integral over $\vec{u}$ can be easily calculated, 
with the result:
\begin{equation}
 \int  \frac{d\vec{u}}{2\,\pi } \;\delta(1-\vec{u}\,^2)\;
 \frac{M_{\alpha}^2}{(u \!\cdot\! P_{\alpha})^2}=  
\int \frac{d\hat{u}}{4\,\pi } \;\; 
\frac{M_{\alpha}^2}{(\hat{u} \!\cdot\! P_{\alpha})^2}
=1\, ,
\label{wf29}
\end{equation}
which is obviously Lorentz invariant. Actually, this property can be 
extended to integrals that have a similar structure. It supposes that 
the extra factor in the integral exhibits the form of a Lorentz scalar 
while its dependence on the 4-vector, $(1,\; \hat{u})$, is invariant 
under the change of scale, 
$(1,\; \hat{u}) \rightarrow  (\lambda, \; \lambda\;\hat{u}) $. 
This property is especially important to 
get Lorentz-invariant expressions for form factors. Using Eq. (\ref{wf29}), 
one recovers the standard normalization for states with different momenta:
\begin{equation}
N(\vec{P}_{\alpha},\vec{P}_{\alpha'})
\equiv E_{P_{\alpha}}\; \delta\Big(\vec{P}_{\alpha}-\vec{P}_{\alpha'}\Big)\;\;
  N^{\alpha \alpha}\;\delta^{\alpha,\alpha'}
= E_{P_{\alpha}}\; \delta\Big(\vec{P}_{\alpha}-\vec{P}_{\alpha'}\Big)\;\;
  N^{\alpha}\;\delta^{\alpha,\alpha'} \, ,
\label{wf31}
\end{equation}
implying $N^{\alpha \alpha}=N^{\alpha}$ for the conventions employed here.
A related expression can be obtained from integrating Eq. (\ref{wf28}) 
over $\vec{P}_{\alpha'}$. It evidences that the normalization 
$N^{\alpha \alpha}$ expressed in terms of the particle momenta does involve 
an integration over the $\vec{u}\; {\rm or}\;(\vec{p}_1+\vec{p}_2)$ 
variable, contrary to other approaches. It is given by:
\begin{eqnarray}
N^{\alpha \alpha} &=& \int \frac{1}{(2\pi)^6}
\frac{d\vec{p}_1}{2\,e_1} \; \frac{d\vec{p}_2}{2\,e_2}\;
\Big(\Phi^{(2)}_{P_{\alpha}}(\vec{p}_1,\vec{p}_2)\Big)^2 \;
8\,\pi^2 \; \delta\Big((p-P_{\alpha})^2\Big)\;M_{\alpha}^2
\nonumber \\
&=& \int \frac{d\vec{k}}{(2\pi)^3}\;(\phi^{\alpha}_0(k))^2  
\int \frac{d\vec{u}}{2\,\pi }\; \delta(1-\vec{u}\,^2)\;
\frac{M_{\alpha}^2}{(u \!\cdot\! P_{\alpha})^2}\, .
\label{wf32}
\end{eqnarray}
%

\noindent
$\bullet$ {\bf Poincar\'e algebra}\\
The general form of the Poincar\'e algebra for the point-form approach 
is known \cite{Keister:sb}. We concentrate here on the 4-momentum 
operator, $P^{\mu}$. Let's remind its general expression:
\begin{equation}
P^0=M\;\sqrt{1+V^2},\;\;\; \vec{P}=M\;\vec{V}\,,
\label{wf33}
\end{equation}
where $\vec{V}$ is the velocity of the system of interest.
For a two-body one, these operators can be obtained from the free-particle
contributions following the Bakamjian-Thomas construction adapted 
to the present case. A change of variables relative to particles 1 
and 2 is first made in terms of the velocity, $\vec{V}$, and the internal 
variable $\vec{k}$:
\begin{eqnarray}
\vec{p_1}_e &=& \vec{k} +\vec{V}\;\frac{\vec{V} \! \cdot \!
\vec{k} }{\sqrt{1+V^2} +1}
 + \vec{V}  \;e_k \, , \;\;
e_{1_e} = \sqrt{1+V^2}\;e_k   + \vec{V}\! \cdot \!  \vec{k}   \, ,
\nonumber \\
\vec{p_2}_e &=& -\vec{k} -\vec{V}\;\frac{\vec{V} \! \cdot \!
\vec{k} }{\sqrt{1+V^2} +1}
 + \vec{V}  \;e_k \, , \;\;
e_{2_e} =  \sqrt{1+V^2} \;\,e_k - \vec{V}\! \cdot \!  \vec{k}   \, ,
\label{wf34}
\end{eqnarray}
with the result:
\begin{equation}
P^0_0=2\,e_k\;\,\sqrt{1+V^2},\;\;\; \vec{P}_0=2\,e_k\;\vec{V}\,.
\label{wf35}
\end{equation}
Anticipating on what follows, a subscript e (for effective) has been 
introduced at the variables relative to particles 1  and 2. 
The second step is to replace the factor $2\,e_k$ by $M$ in 
Eq. (\ref{wf35}), which can be done without modifying the algebra 
provided that the interaction fulfills standard constraints. 

At first sight, the expression of $\vec{V}$ in terms of the single-particle 
momenta can be obtained from Eq. (\ref{wf34}), with the result 
$\vec{V}=(\vec{p_1}_e+\vec{p_2}_e)/(2\,e_k)$. However, it is not a priori 
guaranteed that the momenta appearing in this expression are the genuine 
ones (appearing in the Lagrangian density). From comparing this expression 
with that one given by Eq. (\ref{wf1}) (3-momentum part), it can 
be easily guessed that the answer is negative. A more 
satisfying expression supposes an extra change of variables, which has 
some relationship with the Lorentz-type transformation given by Eq. 
(\ref{wf16}):
\begin{equation}
\vec{p_i}_e = \vec{p_i} +\vec{t}\;\frac{\vec{t} \! \cdot \!
\vec{p_i} }{\sqrt{1+t^2} +1}
 - \vec{t}  \;\,e_i \, , \;\;
e_{i_e} =\sqrt{1+t^2}\; e_i- \vec{t}\! \cdot \!  \vec{p_i}  \;\;(i=1,2) \, ,
\label{wf36}
\end{equation}
with 
\begin{eqnarray}
\vec{t}&=& \frac{2\,e_k}{M}\;\Bigg( \vec{u}  \;
 \frac{ 4\,e_k^2-M^2 }{2\,u \! \cdot \!P}  -
 \vec{p}\;(1-\frac{M}{2\,e_k}) \Bigg)\;X,
\nonumber \\ 
X&=&\frac{2\,e_k\;E_P+e\;M}{e_k\;\Big(4\,e\;E_P-
 (2\,e_k-M)^2 \Big)  }\;\;
\Big(=(e-\frac{\vec{t} \! \cdot \!\vec{p}\,'}{\sqrt{1+t^2} +1})^{-1}\Big)\, ,
\label{wf37}
\end{eqnarray}
where the shortened notations, $e=e_1+e_2$, 
$\vec{p} =\vec{p}_1 +\vec{p}_2$, and 
$\vec{p}-\vec{P}=\vec{u}\;(e-E_P)$, have been used again. 
The sum of the 4-momenta is given by:
\begin{eqnarray}
&&\vec{p}_e=\vec{p}-\vec{t}\;
\Big(e- \frac{\vec{t} \! \cdot \!\vec{p} }{\sqrt{1+t^2} +1}\Big)
=\frac{2\,e_k}{M} \;
\Big(\vec{p}- \vec{u}\; \frac{ 4\,e_k^2-M^2 }{2\,u \! \cdot \!P}\Big)\,,
\nonumber \\
&&e_e=\sqrt{1+t^2}\;\,e- \vec{t}\! \cdot \!  \vec{p}
=\frac{2\,e_k}{M} \;\Big(e- \frac{ 4\,e_k^2-M^2 }{2\,u \! \cdot \!P} \Big)\,.
\label{wf38}
\end{eqnarray}
The above equations may look artificial as they only involve some rewriting 
of previous ones. Their relevance becomes clearer when a comparison of their 
right hand side is made 
with the expression of the 4-momentum $P^{\mu}$, Eq. (\ref{wf1}).
Apart from a factor $2\,e_k/M$, the similarity of these results is complete 
provided that the single-particle momenta, $p_i$, are identified with 
the Lagrangian ones and that the factor $(4\,e_k^2-M^2)/(2\,u \! \cdot \!P)$  
is replaced by the interaction as expected from the mass equation, 
Eq. (\ref{wf21}). This allows one to identify the velocity vector as:
\begin{equation}
\vec{V}= \frac{1}{M} \;
\Big(\vec{p}+ \vec{u}\; \frac{4\,m\;\tilde{V}}{2\,u \! \cdot 
\!P}\Big)\, .
\label{wf39}
\end{equation}
We notice that the Lorentz-type transformation given by Eq. (\ref{wf16}), 
apart from an expected sign difference, is simpler than the one considered 
here. The reason is that the transformation performed in the former case makes 
a term to appear that exactly cancels the interaction one,  leaving as a net 
result a quantity that can be identified to the total 4-momentum, without any 
renormalization. Instead, the transformation given by Eqs. (\ref{wf36}, 
\ref{wf37}), made in the context of the Bakamjian-Thomas construction, 
is less natural, requiring that the factor $2\,e_k$ appearing in Eq. 
(\ref{wf35}) be changed into the mass operator, $M$, by introducing  
the interaction. The difference mainly involves terms of the order
$(1-M/(2\,e_k))$ (apart from expected $\vec{V}$ terms).

\section{Comparison with other approaches}
In this section, we compare different features pertinent to the implementation 
of various forms of relativistic quantum mechanics. They concern the change of 
variables that allows one to show the Lorentz invariance of the mass spectrum 
and the  condition that the interaction has to fulfill in this order. As the 
work was motivated by the peculiar behavior of form factors obtained in an 
earlier implementation of the point form approach, we will consider more 
specifically some aspects of the present implementation with this respect. 
We in particular show how the expected asymptotic behavior is recovered. 

\noindent
$\bullet$ {\bf Change of variables in different forms}\\
When it is compared to other forms, the present implementation of the 
point form evidences a close relationship as for the change of the 
physical variables, $\vec{p}_i$, to the total momentum and the internal 
variable, $\vec{k}$. The transformation is essential to make use of the 
solutions of a mass operator in order, for instance, to calculate 
form factors in different approaches. Interestingly, it is given in all 
cases by a unique expression which generalizes Eq. (\ref{wf18}). 
It now involves a 3-vector $\vec{w}$ that takes the following form: 
\begin{eqnarray}
 \vec{w} =\frac{ \vec{P} }{ 2\,e_k }
+ \frac{ \vec{\xi} }{ 2\,e_k } \; \frac{4\,e_k^2-M^2}{
 \sqrt{ (\xi \cdot P)^2 + \xi^2 \; (4\,e_k^2-M^2)} +  \xi \cdot P} ,
\nonumber \\
\sqrt{1+\vec{w}^2} = \frac{E_P }{ 2\,e_k} +
\frac{\xi^0 }{ 2\,e_k } \frac{4\,e_k^2-M^2 }{ 
 \sqrt{ (\xi \cdot P)^2 + \xi^2 \; (4\,e_k^2-M^2) } + \xi \cdot P } .
\nonumber \\
\label{comp1} 
\end{eqnarray}
The 4-vector $\xi^{\mu}$ appearing in the above expression is specific 
of each approach. Its choice reflects the symmetry properties of the 
hypersurface which the physics is described on. It is uniquely defined 
within each approach and independent of the physical system under 
consideration. As can be noticed, the 4-vector $\xi^{\mu}$ multiplies 
a term $(4\,e_k^2-M^2)$ that can be cast into an interaction one, 
employing Eq. (\ref{wf23}). This is in 
accordance with the expectation that changing the hypersurface 
implies the dynamics. Apart from these features, it can be seen 
that the above expression is independent of the scale of the 4-vector 
$\xi^{\mu}$. Thus, up to an irrelevant scale,  the 4-vector $\xi^{\mu}$ 
is given as follows:\\
- {\it instant form}\\
\begin{equation}
\xi^0=1,\;\;\;  \;\vec{\xi}=0\,, 
\label{comp2} 
\end{equation}
- {\it front form}\\
\begin{equation}
\xi^0=1, \;\;\;  \;\vec{\xi}=\vec{n}\,,
\label{comp3} 
\end{equation}
where $\vec{n}$ is a unit vector with a fixed direction ($\xi^2=0$),\\
- {\it present point form}\\
\begin{equation}
\xi^0=1,  \;\;\;  \;\vec{\xi}=\vec{u}\,,
\label{comp4} 
\end{equation}
where $\vec{u}$ is a unit vector that can point to any direction ($\xi^2=0$). 

To get an invariant mass spectrum in the present approach, it has 
been found that conditions on the interaction have to be fulfilled. 
They are quite similar to those required in the instant- and front-form 
approaches as the consideration of a simple one-boson exchange 
interaction model shows. These conditions are a necessary ingredient. 
They have a close relationship with the fact that higher order terms in the 
interaction have to be included to recover the above invariance property 
when a field-theory approach, based on time-ordered diagrams for instance, 
is used.

\noindent
$\bullet$ {\bf Comparison with an earlier implementation of the point form}\\
The present implementation of the point-form approach differs from 
an earlier one with the above two respects: constraint on the 
interaction and choice of $\xi^{\mu}$. In this last approach,
the relation of the physical variables, $\vec{p}_i$, to the total 
momentum and the internal one assumes the same form as Eq. (\ref{comp1}). 
However, contrary to the present case, the 4-vector $\xi^{\mu}$ is 
determined by the properties of the system under consideration, mainly 
its velocity \cite{Sokolov:1985jv}. 
The corresponding relation, $\xi^{\mu}\propto (P^{\mu}/M)$, 
allows one to considerably simplify Eq. (\ref{comp1}), which then reads: 
\begin{equation}
 \vec{w} =\frac{ \vec{P} }{ M }\, , \;\;\;
\sqrt{1+\vec{w}^2} = \frac{E_P }{ M} \, .
\label{comp5} 
\end{equation}

A first consequence, already mentioned in the introduction, is that the 
kinematical character of the boost transformation is somewhat trivial, 
since both the system and the frame are affected. As a result, 
the constraint that the interaction has to fulfill is much weaker than 
in the other forms, including the implementation of the point form 
presented in this work. A Lorentz invariant one-boson exchange interaction 
is sufficient and no higher order contribution is required to get a Lorentz 
invariant mass spectrum. This can be checked by repeating the different 
steps that lead to the derivation of a mass operator, Eq. (\ref{wf23}), 
or by looking at Eqs. (39, 40) of Ref. \cite{Desplanques:2003nk}, 
where this property is explicitly used. This result suggests that this 
``point form" is not on the same footing as the other forms or the present one.

A second consequence concerns applications of the formalism such as 
calculations of form factors. Implying initial and final states 
with different velocities, and therefore different $\xi^{\mu}$, 
it turns out that these states are described on hypersurfaces defined 
differently. Nothing prevents one from proceeding that way but 
it cannot be considered as the most convenient one. Actually, 
the approach does not account for interaction effects which 
are required so that physics can be described on a hypersurface 
uniquely defined and independent of the physical system under 
consideration. Indeed, Eq. (\ref{comp5}) can be recovered by 
replacing $2\,e_k$ by $M$ in Eq. (\ref{comp1}), which amounts 
to discard the interaction. Including only the boost effect 
common to all other approaches but without the above minimal 
consistency requirements, this approach rather likes a pre-Dirac one.

\begin{figure}[htb]
\begin{center}
\mbox{ \psfig{ file=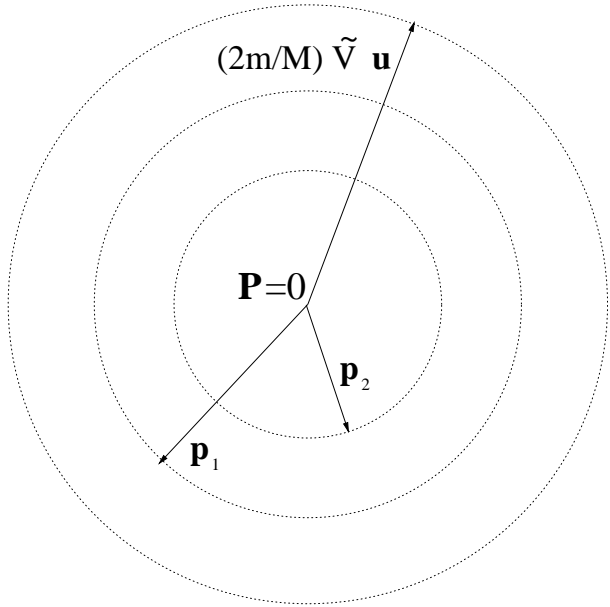, width=8cm}}
\end{center}
\caption{Graphical representation of a configuration representing 
contributions to the total momentum for a system at rest due to momenta, 
$\vec{p}_1$ and $\vec{p}_2$, of particles 1 and 2,  and to the interaction 
part, $(2\,m/M)\;\tilde{V}\; \vec{u}$. It is reminded that the two 
contributions, which sum up here to zero, separately cancel in an earlier 
implementation of the point-form approach  
\cite{Sokolov:1985jv,Lev:1993,Klink:1998}. Contrary to the front-form 
approach, the interaction part and, consequently, the sum of the particle 
momenta, $\vec{p}_1+\vec{p}_2$, points isotropically to all directions, 
which is sketched by circles for the momentum configuration drawn 
in the figure. 
\label{fig1} }
\end{figure}

An observation related to the above remarks concerns the expectation 
value of the 4-momentum operator. In the earlier implementation of the 
point-form approach, this quantity is given by 
\begin{equation}
P^{\mu}=M\;<|\frac{p^{\mu}}{2\,e_k}|>\, ,
\label{comp6}
\end{equation}
which is the simplest possible choice. It assumes that the 
constituent-momentum and interaction parts are separately proportional 
to the velocity of the system. A zero 3-momentum value (c.m. case)
assumes that $\vec{p}=\vec{p}_1+\vec{p}_2=0$. Such a result holds more 
generally; it however involves quasi particles rather than physical ones. 
When considering these last degrees of freedom, the present implementation 
evidences a different pattern, perhaps more complicated but more 
consistent with describing physics on a unique hypersurface. 
A zero value can be obtained by averaging the non-zero 3-vector, 
$\vec{p}_1+\vec{p}_2$, over all directions but it can also be checked  
that this single-particle contribution, for some direction, is cancelled 
by an interaction part, accordingly to Eq. (\ref{wf39}). This 
example provides a nice illustration of the dynamical character of 
the momentum operator in the point-form approach. A graphical 
representation for a system at rest is given in Fig. \ref{fig1}.
More generally, the sum of the constituent momenta fluctuates around the total
momentum with an amplitude determined by the interaction\footnote{To some extent, there is a similarity with the 
"Zitterbewegung" effect, but in momentum rather than in configuration space. 
We are grateful to Ica Stancu for proposing this analogy.}.

\noindent
$\bullet$ {\bf Asymptotic behavior of form factors}\\
Form factors calculated in the earlier implementation of the point-form 
approach show a fall off faster than expected. For the ground state 
of a system made of scalar constituents, the asymptotic behavior is 
$Q^{-8}$, instead of $Q^{-4}$ \cite{Desplanques:2001zw,Amghar:2002jx}. 
The difference can be traced back to a peculiarity of the formalism  
that changes  the dependence on a factor $Q^2$ into a dependence 
on $Q^2\,(1+Q^2/(4\,M^2))$ \cite{Allen:2000ge}. A fast fall off also 
appears in the case where the mass of the system is small compared 
to the sum of the constituent masses (the pion for instance) with 
the result that the charge radius scales like the inverse of the 
mass \cite{Amghar:2003tx}. The question arises whether the present 
implementation of the point form solves these problems.

\begin{figure}[htb]
\begin{center}
\mbox{ \psfig{ file=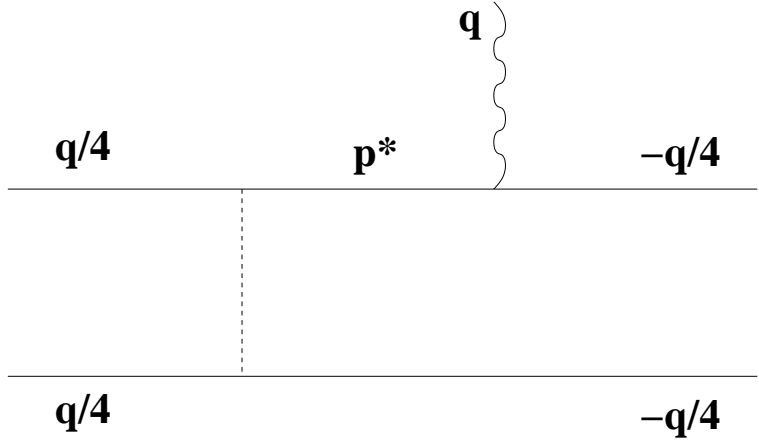, width=10cm}}
\end{center}
\caption{Virtual scalar particle or photon absorption on a two-body system 
in the Born approximation. A similar diagram with a different time ordering 
of the boson exchange and the interaction with the external probe should be 
considered. The kinematical definitions refer to the Breit-frame and an 
infinitesimally small binding. The momentum of the intermediate particle, 
$\vec{p}\,^*$, depends on the form of relativistic quantum mechanics. 
\label{fig2} }
\end{figure}

Detailed calculations of form factors along the present approach will be 
presented elsewhere. We only show here, schematically, how the approach 
allows one to get the expected power law $Q^{-4}$ for the scalar 
constituent case (ground state). At high $Q^2$, the form factor is
given by the Born amplitude whose graphical representation is given 
in Fig. \ref{fig2} for the Breit-frame configuration and in the limit 
of small binding. In order to get an estimate, the value taken 
by the momentum of the intermediate particle, $\vec{p}\,^*$, has to be 
determined. The  amplitude thus depends on the relation that implies 
this momentum, the spectator one and the total momentum and, therefore, 
on the form of relativistic quantum mechanics under consideration. 

For the present point-form approach, the relation (\ref{int1}) reads:
\begin{equation} 
|\vec{p}\,^*-\frac{1}{4} \; \vec{q}-\frac{1}{2} \; \vec{q}|
= |e^*+e_{Q/4} -2\,e_{Q/4}|.
\label{comp7}
\end{equation}
Writing $\vec{p}\,^*=\alpha\; \vec{q}/4$, one gets in the high 
$Q$ limit the equation $|\alpha-3|\; Q= |\alpha-1|\;Q$, which implies 
$\alpha \rightarrow \infty$ (no real interest here) and $\alpha= 2$. 
Solutions with a component of $\vec{p}\,^*$ perpendicular to $\vec{q}$ 
are also possible (for instance $\alpha= 3$ and a component 
of $\vec{p}\,^*$ perpendicular to $\vec{q}$ equal to 4 in units of $Q/4$, 
which corresponds to the front-form case with $q^+=0$).

In the earlier ``point-form" approach, the equation to be fulfilled 
is given by: 
\begin{equation} 
\vec{p}\,^* - \frac{\vec{q}}{4}= (e^*+e_{Q/4})\;
\frac{\vec{q}/2}{2\,e_{Q/4}}.
\label{comp8}
\end{equation}
The above equation is verified with $\alpha=3+Q^2/(4\,m^2)$, which 
contrary to the above case or the instant-form one ($\alpha=3$), tends 
to $\infty$  when $Q\rightarrow \infty$. To determine how the Born amplitude 
depends on  the choice of the form, one has to insert  the appropriate 
expression of the $\vec{k}$ variable in the factor that results from 
the meson and the two-body propagator in Fig. (\ref{fig2}). In the limit 
of large momentum transfers, one thus obtains:
\begin{equation} 
\frac{g^2}{\mu^2+(\vec{k}-\vec{k}\,')^2}\;\;\frac{1}{4\,e_k^2-M^2} \simeq
\frac{32\,g^2}{2\,\alpha\;( Q^2+4\,m^2)} \;\; 
\frac{8}{2\,\alpha\;( Q^2+4\,m^2)}\, .
\label{comp9}
\end{equation}
Replacing $\alpha$ by its expression, one gets the following asymptotic 
behaviors:
\begin{eqnarray}
&&16\,g^2\;Q^{-4}\;\;({\rm present\;\; point\;\;  form}) \, ,
\nonumber \\
&&4\,g^2\;\frac{Q^{-4}}{\Big(1+Q^2/(16\,m^2)\Big)^2} \propto Q^{-8}
\;\;({\rm earlier\;\;  ``point\;\;  form"})\, .
\label{comp10}
\end{eqnarray}
It can be  seen that the difference between the two results at high $Q^2$ 
is  essentially due to a factor $(1+Q^2/(16\,m^2))$ that multiplies each 
$Q^2$ term\footnote{In spite of approximations in Eq. (\ref{comp9}), the factor 
is an exact consequence of Eq. (\ref{comp8}).}. As can be checked in the 
small binding limit, this factor is identical to the one already found 
in \cite{Allen:2000ge}, reminded at the beginning of this subsection. 
It is noticed that the front-form case with $q^+=0$ gives the asymptotic 
behavior $4\,g^2\;Q^{-4}$, in agreement with the exact result, while the
instant-form one gives an extra factor $16/9$, requiring in any case two-body
currents to get it.

\section{Conclusion} 
In this work devoted to the description of a two-body system within  
relativistic quantum mechanics, we considered a point-form approach 
inspired by the Dirac's one, which implies that physics is described 
on a hyperboloid surface. The main difficulty resides in solving 
the corresponding wave equation, 
taking into account that the integration of plane wanes over a hyperboloid 
surface provides expressions that are not so easy to deal with as the 
3-dimensional $\delta(\cdots)$ functions obtained for a hyperplane. 
For the solution we were able to find, which corresponds to a particular case
considered by Dirac, the relation of the constituent 
momenta to the total momentum of the two-body system we are interested in
is given by the equation:    
$$\vec{p}_1+\vec{p}_2-\vec{P}-\vec{u}\;(e_1+e_2- E_P)=0,$$ 
where  
$\vec{u}$ is a unit vector whose direction is to be integrated over. 
Provided that the interaction has the appropriate properties pertinent 
to the construction of the Poincar\'e algebra, it turns out that this 
direction is conserved by the interaction. This property greatly 
simplifies the search for a solution. At the same time, it allows 
one to establish some relationship with the front-form approach 
where a similar relation holds but for a fixed direction (defined 
by the orientation of the hyperplance which physics is described on). 
In view of an expected unitary equivalence between different forms of 
relativistic quantum mechanics, this relationship is not surprising.

The most striking feature evidenced by the approach arises from 
the above relation of the constituent momenta to the total momentum 
of the system of interest. The sum of these constituent momenta 
at a given $\vec{P}$, contrary to other forms, is not a fixed quantity. 
In the c.m. frame, it differs from zero off-energy shell and, moreover, 
points isotropically to all spatial directions, in agreement with 
the absence of preferred direction on the hyperboloid surface. 
Despite the unfamiliar character of this approach, it nevertheless 
appears that a consistent scheme could be developped. In particular, 
the correct total momentum is recovered when the contribution to it 
due to the interaction is accounted for, as expected from the dynamical 
character of the 4-momentum in the point-form approach.

Concerning other aspects of the present implementation of the point-form 
approach, it appears that the relation of the physical momenta to the 
total momentum and an internal variable is formally quite similar 
to that for the instant- and front-form approaches. The only difference 
is in the choice of a 4-vector which characterizes the hypersurface 
which physics is described on and reduces to the 4-vector $(1,\;\vec{u})$ 
in the present case. The derivation of the mass operator requires 
the same kind of constraints than in the other forms. The standard 
expression of the normalization in terms of the solution of the 
mass operator is recovered, with  a difference in its derivation however. 
It implies a further integration over the orientation of the total momentum 
carried by the constituents, $\vec{u}$. 

The present work was motivated for a part by the drawbacks evidenced 
by the form factors calculated in an earlier ``point-form" implementation
\cite{Desplanques:2001zw,Amghar:2002jx}. 
In comparison to an ``exact" calculation, the fall off was too fast. 
For the ground state of a system made of scalar constituents, 
the asymptotic behavior was rather given by the power law $Q^{-2n}$ 
where $Q^{-n}$ is expected. The slope at small momentum transfers  
was scaling like $M^{-2}$, providing a squared charge radius that tends 
to $\infty$ when the mass tends to zero! As this ``point-form" 
implementation was implying hyperplanes, moreover different for the 
initial and final states in the case of form factors, the question 
arises of what this approach is responsible for in the peculiar
features evidenced by these form factors. The comparison with results 
obtained in the present point-form implementation, based on describing 
the physics on a hyperboloid, suggests that the first problem 
is specific of the earlier ``point-form" implementation. As worked 
out in Ref. \cite{Desplanques:2003nk}, this problem could be cured 
by considering specific two-body currents. The other 
problem, related to the dependence of form factors on the momentum 
transfer $Q$ through the quantity  $Q/2M$ in both cases, remains complete.
Up to recently, it could be thought that this second problem was 
a characteristic feature of the point-form approach. Examination 
of a few not well known results obtained in the light-front approach 
for a frame $q^+\neq0$ \cite{Bakker:2000pk,Simula:2002vm,deMelo:2002yq} or in 
the instant-form approach for a parallel kinematics \cite{Amghar:2002jx} 
shows similar features when a one-body current is retained. This 
relationship clearly points to an important role of missing two-body 
currents, which have been identified in some field-theory based 
calculations \cite{Bakker:2000pk,Simula:2002vm,deMelo:2002yq,Amghar:2002jx} 
but are different from those considered in Ref. \cite{Desplanques:2003nk}. 
Thus, the present point-form implementation solves one of the problem 
raised by an earlier one, namely the wrong power-law behavior of 
the asymptotic form factor. With this respect, it now compares to the 
other forms of relativistic quantum mechanics. The other problem, 
illustrated by the scaling of the charge radius with the inverse of the 
mass of the system, appears to be a more general one. It probably points 
to a sizeable violation of Poincar\'e space-time translation invariance in
calculating form factors  \cite{Desplanques:2004}.

As the present and the earlier implementations of the point-form 
approach satisfy the main properties pertinent to the related Poincar\'e 
algebra, one may wonder why they differ in their predictions. 
Looking for some genuine explanation, it appears that the difference 
resides in how the velocity vector, $\vec{V}$, is related to the physical 
degrees of freedom. In the simplest case and for a two-body system, 
$\vec{V}\propto (\vec{p}_1+\vec{p}_2)$, which is reminiscent of the free 
particle system. However, these momenta could have an effective character 
and have a more complicated expression in terms of the physical momenta 
and the interaction, quite consistently with the dynamical character 
of the momentum in the point-form approach. To some extent, this offers 
some freedom that is used to fulfill the requirement that physics 
be described on the same hypersurface, whatever the system under 
consideration. Such a property could not be obtained in the simplest case.

While considering equations with physics described on a hyperboloid, 
the form of the solution we found was partly guessed on the basis 
of minimal symmetry and consistency requirements. On the other hand, 
it sounds that the calculation of form factors in the present 
point-form implementation amounts to make an appropriately 
weighted average of the front-form form factor over the orientation 
of the front. In some sense, this relationship is fortunate 
and can provide useful hints. However, one can wonder why the 
point-form calculation of form factors should reduce to this simple 
recipe. This feature being closely related to the solution we found, 
the question arises whether this is the more general one. Finding 
an answer is a task for the future. For the time being, we believe 
that the present solution evidences features different enough from 
what other forms of relativistic quantum mechanics show and, in this 
respect, represents a stimulating starting point for further research. 

{\bf Acknowledgements} \\
We are very grateful to S. Noguera for interesting and useful comments 
concerning the presentation of this work. We would also like to acknowledge 
A. Amghar and L. Theu{\ss}l for preliminary studies or discussions 
at the earlier stage of this work.

\appendix

\section{Dealing with a hyperboloid surface: limit $\tilde{\tau}=0$} 
\label{app:a}
We here consider expressions involving the integration of plane waves 
over a hyperboloid surface,  $x^2-\tilde{\tau}=0$. This corresponds to 
$\tilde{\tau}>0$ but, actually, the part with $\tilde{\tau}<0$, outside the
light cone, should be considered for completeness. Moreover, the forward and
backward ``times'' have to be included in the discussion. As explained 
in the text, we concentrate on the particular case $\tilde{\tau}=0$, which is 
considerably simpler \cite{Dirac:1949cp}. For illustration, we consider 
a minimal but essential application, and show how the contribution of 
a free scalar particle to the 4-momentum is recovered. To allow for 
a comparison, some results for the case of a hyperplane are first reminded. 

The contribution of free scalar particles to the 4-momentum may be 
generally written as:
\begin{equation}
``P_0^{\mu}\,"= \int d^4x \; f_{\nu}(x) \;
\partial^{\nu}\phi(x) \;\partial^{\mu}\phi(x) \, , 
\label{appa1}
\end{equation} 
where $f_{\nu}(x)$ characterizes the surface which the integral is performed 
over. Keeping the only term of interest here, one gets:
\begin{eqnarray}
&& ``P_0^{\mu}\,"=
\int \frac{ d\vec{p}\;\;d\vec{p}\,'}{(e_p\;e_{p'})^{1/2}\;(2\,\pi)^3} 
\; a^*(\vec{p}) \;a(\vec{p}\,')\; p^{\mu} \; p'^{\nu} \;I_{\nu}, 
\nonumber \\
&&{\rm with} \;\;\; I_{\nu} =\int d^4x \; f_{\nu}(x) \;e^{i(p-p')\cdot x} \, . 
\label{appa2}
\end{eqnarray} 
Anticipating on a generalization of the hyperplane case to different 
hypersurfaces, the function $f_{\nu}(x)$ may be written as the limit for 
$\tilde{\tau}\rightarrow 0$ of the function:
\begin{eqnarray}
 f^{\tilde{\tau}}_{\nu}(x)&=& -\partial_{\nu}\;\Theta(x,\tilde{\tau})
  \nonumber \\   {\rm with}\;\;
\Theta(x,\tilde{\tau})&=& 
  \theta( \xi \!\cdot\! x) \; \theta\Big(\tilde{\tau}-F(x) \Big)
+  \theta(-\xi \!\cdot\! x) \; \theta\Big(\tilde{\tau}+F(x) \Big)\, ,
\label{appa3}
\end{eqnarray} 
where $F(x)$ and $\xi^{\mu}$ are arbitrary at this point but should be chosen 
accordingly to the hypersurface of interest.

The above $\Theta(x,\tilde{\tau}) $ function\footnote{The notations 
$\tilde{\tau}$ and $\tau$ introduced below differ in that they 
respectively involve constraints chosen here as quadratic and linear 
in the $x$ variable.}  determines a partition of space
(0 for $\tilde{\tau}=-\infty$ and 1 for $\tilde{\tau}=\infty$). 
Its introduction does not modify the derivation of the 4-momentum 
conservation in a simple case which therefore represents a minimal 
requirement  whereas its derivative with respect to $\tilde{\tau}$ (or $x$) 
defines the hypersurface of interest in term of the $\tilde{\tau}$ variable. 
This is illustrated by the following relations:
\begin{eqnarray}
&& \hspace{-5mm}(2\,\pi)^4 \; \delta^4(p-p') =\int d^4x  \;e^{i(p-p')\cdot x}
 = \int d^4x \;e^{i(p-p')\cdot x} \int_{-\infty}^{\infty} d\tilde{\tau}\;
\Bigg(\frac{d}{d\tilde{\tau}}\Theta(x,\tilde{\tau})\Bigg) 
 \nonumber \\   && \hspace{5mm}
=\int_{-\infty}^{\infty} d\tilde{\tau} \int d^4x \;e^{i(p-p')\cdot x}\;
\Bigg( \theta( \xi \!\cdot\! x)\;
\delta\Big(\tilde{\tau}-F(x) \Big)
+ \theta( -\xi \!\cdot\! x)\;
\delta\Big(\tilde{\tau}+F(x) \Big) \Bigg)\, . 
\label{appa4}
\end{eqnarray} 
The hypersurface defined by the last factor at the r.h.s. of this equation must
be such that the whole space-time volume be visited when $\tilde{\tau}$ 
varies from $-\infty$ to $+\infty$.

\noindent
$\bullet$ {\bf Hyperplane case}\\
For a hyperplane with orientation $\lambda_{\nu}$, the function 
$f^{\tau}_{\nu}(x)$ may be written as:
\begin{equation}
 f^{\tau}_{\nu}(x) =- \partial_{\nu}\; \theta(\tau-\lambda \!\cdot\! x)
 =\lambda_{\nu}\; \delta(\lambda \!\cdot\! x - \tau)  \, , 
\label{appa5}
\end{equation} 
corresponding to take in Eq. (\ref{appa3}) $ F(x)=(\lambda \!\cdot\! x)^2$ 
and $\xi^{\mu}=\lambda^{\mu}$. 

It is noticed that the expression of $f^{\tau}_{\nu}(x)$  at $\tau=0$ is 
unchanged when  $x\rightarrow -x$. Using the corresponding result of the 
integral over the variable $x$: 
\begin{eqnarray}
I_{\nu}=\int d^4x  \;\lambda_{\nu}\;\delta(\lambda \!\cdot\! x )\; 
e^{i(p-p')\cdot x}
&=&\frac{\lambda_{\nu}}{\lambda^0}\; (2\,\pi)^3 \; \delta\Big(\vec{p}-\vec{p}\,'
-\frac{\vec{\lambda}}{\lambda^0}\;(e_p-e_{p'})\Big)
\nonumber \\ &=& \lambda_{\nu}\;(2\,\pi)^3 \;
\frac{e_{p}\;\delta(\vec{p}-\vec{p}\,') }{\lambda \cdot p} \, ,
\label{appa6}
\end{eqnarray} 
one gets the desired expression: 
\begin{equation}
``P_0^{\mu}\,"=\int d\vec{p} \; a^*(\vec{p}) \; a(\vec{p})\;  p^{\mu} \; \, . 
\label{appa7}
\end{equation} 
As expected from Eq. (\ref{appa5}) for $\tau=0$, the result is independent 
of the scale of the 4-vector, $\lambda^{\mu}$. It does not depend either 
on its direction. 

\noindent
$\bullet$ {\bf Hyperboloid case}\\
In dealing with a hyperboloid surface, we start from  Eq. (\ref{appa3}) 
which, as previously reminded, is consistent with minimal requirements. 
Disregarding a possible constant term, the function $F(x)$ that appears 
there is now taken as $F(x)=x^2$ while 
the general 4-vector $\xi^{\mu}$, essential to distinguish between backward and 
forward times, is denoted $U^{\mu}$ with $U^2 \geq 0$.  
The  function $f^{\tilde{\tau}}_{\nu}(x)$ thus obtained reads:
\begin{equation}
 f^{\tilde{\tau}}_{\nu}(x)= 2\, x_{\nu}\; 
\Big( \theta( U \!\cdot\! x)\;
\delta(\tilde{\tau}-x^2 )
- \theta( -U \!\cdot\! x)\;
\delta(\tilde{\tau}+x^2 ) \Big)+ \dots \, ,
\label{appa9}
\end{equation} 
where dots represent a contribution irrelevant for our purpose. 
In considering the limit of the above expression for $\tilde{\tau}=0$, it is 
appropriate to rewrite it as follows:
\begin{eqnarray}
 f^{\tilde{\tau}}_{\nu}(x)=  x_{\nu} 
&&\hspace{-6mm} \Bigg(
\Big( \delta(\tilde{\tau}-x^2 )+\delta(\tilde{\tau}+x^2 )\Big) \;
\Big[ \theta( U \!\cdot\! x)- \theta( -U \!\cdot\! x)\Big]
\nonumber \\
&&+\Big( \delta(\tilde{\tau}-x^2 )-\delta(\tilde{\tau}+x^2 )\Big) \;
\Big[ \theta( U \!\cdot\! x)+ \theta( -U \!\cdot\! x)\Big] \Bigg)+ \dots\, .
\label{appa10}
\end{eqnarray} 
Due to the symmetrical character of the integration on positive and negative 
values of $\tilde{\tau}$, the last term as well as the dots part can be ignored. Concerning the first 
term, we notice that the two $\delta(\cdots)$ functions exclude each other. 
When taking the limit $\tilde{\tau} \rightarrow 0$, where the hyperboloid 
defined by the equation $\tilde{\tau}-x^2=0$ reduces to the light-cone, 
only one of these terms should be therefore retained. The above expression 
then simplifies to read: 
\begin{equation}
 f_{\nu}(x)=  x_{\nu}\;\delta(x^2)\;
\Big[\theta( U \!\cdot\! x)- \theta(- U \!\cdot\! x) \Big]. 
\label{appa11}
\end{equation} 
The last factor may look surprising but, together with the front factor 
$x_{\nu}$, it simply accounts for the space-time symmetry, 
$x \rightarrow -x$, and in particular for the symmetry between backward 
and forward times in the case $\nu=0$ and $U^{\mu}= 1,\;0,\;0,\;0$.

When considering the contribution to the total momentum at $\tilde{\tau}=0 $, 
expressions given by Eq. (\ref{appa2}) have again to be considered. 
The quantity $I_{\nu}$, which now involves the expression of $ f_{\nu}(x)$ 
given by Eq. (\ref{appa11}), now reads:
\begin{eqnarray}
I_{\nu}&=&\int d^4x  \;x_{\nu}\;\delta( x^2 )\;
\Big[ \theta(U\!\cdot\! x)- \theta(-U\!\cdot\! x) \Big] \; e^{i(p-p')\cdot x}
\nonumber \\ 
&=&-i\;\partial_{(p-p')^{\nu}} \Bigg( \int d^4x  \;   \delta(x^2)  \;  
\Big(\theta(U \!\cdot\! x)-  \theta(-U \!\cdot\! x) \Big) \;
\;e^{i\, (p-p')\cdot x}\Bigg)
\nonumber \\ 
&=&4\,\pi^2\partial_{(p-p')^{\nu}}\;  
\Bigg( \delta\Big((p-p')^2\Big)\;\theta\Big(U \!\cdot\! (p-p')\Big)  
- \delta\Big((p-p')^2\Big)\;\theta\Big(U \!\cdot\! (p'-p) \Big)  \Bigg)
\nonumber \\ 
&=& 8\,\pi^2\; U_{\nu}\;\delta\Big((p-p')^2\Big) \; 
\delta\Big(U \!\cdot\! (p-p') \Big)
\nonumber \\ && \hspace{1cm} 
 +8\,\pi^2\;(p-p')_{\nu} \; \delta'\Big((p-p')^2\Big) \;
\Big[\,\theta\Big(U \!\cdot\! (p-p')\Big)
- \theta\Big(U \!\cdot\! (p'-p)  )\Big)\,\Big].
\label{appa12}
\end{eqnarray} 
The above result can be extended without any change to a many-body system. 
Steps have been detailed as some of them should be considered with caution.  
It has been assumed that the 4-vector $U^{\mu}$ was independent of $p-p'$. 
This may be the case for the present application. 
However, for other ones considered in this work, this will not be always the 
case. We checked that, for those applications, the corresponding contribution 
contains a factor which vanishes due to the presence of a $\delta(\cdots)$ 
function. What matters is that the relevant dependence on $p-p'$, in the 
$\theta(U \!\cdot\! (p-p') ) $ function for instance, is already factored out. 
On the other hand, for the application considered here,  
the very last term in Eq. (\ref{appa12}) ($8\,\pi^2 \cdots$) cannot 
contribute since it contains the factor $p-p'$ 
that should ultimately vanish (conservation of the 4-momentum). As to 
the other term, $ \delta(U \!\cdot\! (p-p') ) \; \delta\Big((p-p')^2\Big)$, 
one could guess that the two $\delta(\cdots)$ functions produce 
$\delta(\cdots)$ functions implying the components of $(p-p')$ along 
respectively some direction and the transverse ones (once the other  
one has been accounted for), then giving rise to the expected  
3-dimensional $\delta(\cdots)$ function, $\delta(\vec{p}-\vec{p}\,')$ 
(see Eq. (\ref{appd2}) of appendix \ref{app:d}).  
One then gets: 
\begin{equation}
I_{\nu} =U_{\nu}\;(2\,\pi)^3 \;
\frac{e_{p}\;\delta(\vec{p}-\vec{p}\,')}{U \cdot p}  \, .
\label{appa13}
\end{equation} 
This result is quite similar to Eq. (\ref{appa6}) obtained for a hyperplane 
surface. It shows how various $\delta(\cdots)$ functions implying scalar 
arguments combine to give a 3-dimensional $\delta(\cdots)$ function. 
Let's finally notice that the combination of the $\theta(\cdots)$ 
in Eq. (\ref{appa11}) allows one to get rid of a term that has the 
wrong imaginary phase and, moreover, no expected role here. 

\section{Getting a solution of the wave equation} 
\label{app:b}
In looking for a solution of a wave equation, Eq. (\ref{wf8}) 
for instance, we have in mind that 
the effect of the interaction should leave unchanged the function, 
$f_{\vec{P}}(\vec{p}_1,\vec{p}_2)$, which relates the constituent 
momenta to the total momentum of the system under consideration. 
In the instant-form case, where the interaction conserves 
the 3-momentum, this is trivially fulfilled by taking 
$f_{\vec{P}}(\vec{p}_1,\vec{p}_2)=\delta(\vec{p}_1+\vec{p}_2-\vec{P})$, 
as can be checked from the equality:
\begin{equation}
\delta(\vec{p}_1+\vec{p}_2-\vec{P})=\int d(\vec{p}\,'_1+\vec{p}\,'_2)\; 
\delta(\vec{p}_1+\vec{p}_2-\vec{p}\,'_1-\vec{p}\,'_2)\;
\delta(\vec{p}\,'_1+\vec{p}\,'_2-\vec{P})\, .
\label{appb1}
\end{equation}
In the present case, the $\delta(\cdots)$ functions appearing 
in the interaction involve some quadratic dependence on 4-momenta, 
see Eq. (\ref{appa12}). They cannot be simply factorized into 
terms involving the components of these 4-momenta separately. 
Thus, the derivation of a relation that would have the structure 
of the above equation is not so trivial. 

\noindent
$\bullet$ {\bf Dealing with Lorentz-scalar $\delta(\cdots)$ functions}\\
The key quantity  that we have to deal with and replaces 
the 3-dimensional $\delta(\cdots)$ function ensuring momentum 
conservation in the above equation is obtained from contracting 
$I_{\nu}$, Eq. (\ref{appa12}), with $(p+p')^{\nu}$ where $p$ now 
represents the sum of the 4-momenta of particles 1 and 2. It is 
given by: 
\begin{eqnarray}
J&=&8\,\pi^2 \Bigg(U \!\cdot\! (p+p')\;\delta\Big((p-p')^2\Big)\;
\delta\Big(U \!\cdot\! (p-p')\Big) 
\nonumber \\ && \hspace{1.5cm} 
+\,(p^2-p'^2) \; \delta'\Big((p-p')^2\Big)
 \Big[ 
\theta\Big(U \!\cdot\! (p-p') \Big)
- \theta\Big(U \!\cdot\! (p'-p) \Big)
\Big] \Bigg)\, . 
\label{appb2} 
\end{eqnarray} 
In order to find a solution to our problem, we assume that the 4-vector U 
can be written as a linear combination of 4-momenta, $p-P$ and $p'-P$,
as in Eq. (\ref{wf9}) . 
Moreover, we assume that the above quantity, $ J$, is multiplied by 
the function  $\delta\Big((p'-P)^2\Big)$, which could be a part of the 
solution we look for. To deal with  Eq. (\ref{appb2}), 
the trick is to insert P terms so that to get an expression involving 
mainly combinations of $p-P$ and $p'-P$ and to cancel part of the arguments 
in the $\delta(\cdots)$ functions taking into account the other ones.
As an example, we give the various steps concerned with the first term 
in Eq. (\ref{appb2}): 
\begin{eqnarray}
&&\delta\Big((p-p')^2\Big)\;\;\delta\Big(U \!\cdot\! (p-p')\Big) \;\;
\delta\Big((p'-P)^2\Big)
\nonumber \\ &&= 
\delta\Big((p-P)^2-2\,(p-P)\!\cdot\!(p'-P)+(p'-P)^2\Big)\;\;
\delta\Big((p'-P)^2\Big)
\nonumber \\
&& \;\;\; \times  \, \delta\Bigg(c\;
\Big((p-P)^2-(p-P)\!\cdot\!(p'-P)\Big)
 -c'\;\Big((p'-P)^2-(p-P)\!\cdot\!(p'-P)\Big)\Bigg)
\nonumber \\ &&= 
\delta\Big((p-P)^2-2\,(p-P)\!\cdot\!(p'-P)\Big)\;\delta\Big((p'-P)^2\Big)
\nonumber \\
&& \;\;\; \times \, \delta\Bigg((p-P)\!\cdot\!(p'-P)\;\;
\Big(c+c'\Big)\;\Bigg)
\nonumber \\ &&
= \delta\Big((p-P)^2\Big)\;\; \delta\Big((p-P) \!\cdot\! (p'-P)\Big) \;\;
 \delta\Big((p'-P)^2\Big)\;\;
\Big( c+c'\Big)^{-1} \,.
\label{appb3} 
\end{eqnarray} 
Using the various $\delta(\cdots)$ functions, the factor multiplying 
the above term in Eq. (\ref{appb2}), $U \!\cdot\! (p+p')$, can be 
transformed as follows:
\begin{equation}
U \!\cdot\! (p+p')= 2\,U \!\cdot\! P= c\,(p^2-M^2)+c'\,(p'^2-M^2)\,.
\label{appb4}
\end{equation}
The second term in Eq. (\ref{appb2}) can be similarly dealt with except 
for the derivative of the $\delta(\cdots)$ function. This one can be 
tranformed away by an integral by parts that changes the $\theta(\cdots)$ 
function into a $\delta(\cdots)$ function. While doing so, one has 
to assume that the remaining part of the integrand does not depend 
on $(p-P) \!\cdot\! (p'-P)$. This condition is no more than the 
constraint that the interaction in the mass operator has to fulfill 
in any case in the various forms of relativistic quantum mechanics. 
Gathering the different results, one obtains:
\begin{eqnarray}
&&\Bigg(U \!\cdot\! (p+p')\;\delta\Big((p-p')^2\Big)\;
\delta\Big(U \!\cdot\! (p-p')\Big) 
\nonumber \\ && \hspace{1.0cm} 
+\,(p^2-p'^2) \; \delta'\Big((p-p')^2\Big)
 \Big[ 
\theta\Big(U \!\cdot\! (p-p')\Big)
- \theta\Big(U \!\cdot\! (p'-p) \Big)
\Big] \Bigg) \; \delta\Big((p'-P)^2\Big) \hspace{0.7cm}
\nonumber \\ && =
\delta\Big((p-P)^2\Big) \; \delta\Big((p-P) \!\cdot\! (p'-P)\Big) \; 
\delta\Big((p'-P)^2\Big)\;
\frac{c\;(p'^2-M^2)+c'\;(p^2-M^2)}{c+c'} \, .
\label{appb5} 
\end{eqnarray} 

\noindent
$\bullet$ {\bf Relation $\vec{u}= \vec{u}\,'$}\\
It is now interesting to consider the consequences implied by the 
simultaneous presence of the three $\delta(\cdots)$ functions 
in the last equation. We first notice that those of them with 
the argument $(p-P)^2$ or $(p'-P)^2$ lead to relations such as:
\begin{equation}
\vec{u}= \frac{\vec{p}-\vec{P}}{e-E_P}\, , \;\;\; 
\vec{u}\,'= \frac{\vec{p}\,'-\vec{P}}{e'-E_P}\, ,
\label{appb6} 
\end{equation}
where $\vec{u}$ and $\vec{u}\,'$ are unit vectors. Considering now 
the second $\delta(\cdots)$ function at the last line of Eq. (\ref{appb5}), 
it is seen that it can be written in a way where the dependence 
on $\vec{u}$ and $\vec{u}\,'$ is factorized:
\begin{eqnarray}
\delta\Big((p-P) \!\cdot\! 
(p'-P)\Big)&=&\delta\Big((1-\vec{u}\!\cdot\!\vec{u}\,')\;
(e-E_P) \;(e'-E_P)\Big)
\nonumber \\
&=&\frac{\delta\Big(1-\vec{u}\!\cdot\!\vec{u}\,'\Big)}{|e-E_P| \;|e'-E_P|}\, .
\label{appb7} 
\end{eqnarray}
As $\vec{u}$ and $\vec{u}\,'$ are unit vectors, this last relation 
implies that their relative angle is zero, hence the relation:
\begin{equation}
\vec{u}= \vec{u}\,'\, .
\label{appb8} 
\end{equation}
It is important to notice that such a result could not be obtained 
if one of $\delta(\cdots)$ function with the argument $(p-P)^2$ 
or $(p'-P)^2$ was absent in Eq. (\ref{appb5}). For practical 
purposes, it supposes some information on the wave function 
($1-\vec{u}\,^2=0$). This feature contrasts with the instant or front 
forms where a 3-dimensional equality also holds (conservation 
of the 3-momentum in the instant form for instance, see 
Eq. (\ref{appb1})). In these cases, the relation occurs at the level 
of the interaction alone, independently of the wave function. 

Another important consequence concerns the 4-vector $U^{\mu}$. 
Its definition in the interaction context, Eq. (\ref{wf9}), was 
involving both the initial and final states. The above result 
makes the definition free of ambiguity, allowing one to identify 
the 4-vectors $U^{\mu}$ and $u^{\mu}$ (up to a factor). On the 
one hand, one has:
\begin{equation}
(e-E_P,\;\vec{p}-\vec{P}) \propto (e'-E_P,\;\vec{p}\,'-\vec{P})
 \propto (1,\;\vec{u}) \propto (U^0,\;\vec{U})\, .
\label{appb9} 
\end{equation}
On the other hand, the proportionality factor does not matter. The 
4-vector $U^{\mu}$ is defined up to a factor in any case and, moreover, 
it has a zero norm, which is essential to get rid of various scales. 

\section{Change of variables} 
\label{app:c}
The Jacobian transformation resulting from the changes of variables,  
Eqs. (\ref{wf16}) and (\ref{wf18}), are successively given by:
\begin{eqnarray}
\int \frac{d\vec{p}_1}{2\,e_1} \; \frac{d\vec{p}_2}{2\,e_2}\;\cdots&&=
\int d\vec{k}\;2\,e_k \frac{ d\vec{w} }{\sqrt{1+w^2}}\;\cdots
\nonumber \\
&&=\int d\vec{k}\;2\,e_k\;  d\vec{u} \;
\frac{2\,e_k}{\sqrt{ (u \!\cdot\! P)^2 +(1-\vec{u}\,^2)\,(4\,e_k^2-M^2 )  }}
\nonumber \\ && \hspace{1cm} \times
\Bigg(\frac{4\,e_k^2-M^2}{2\,e_k\;
\Big(u \!\cdot\! P+ \sqrt{ (u \!\cdot\! P)^2 
+(1-\vec{u}\,^2)\,(4\,e_k^2-M^2 ) } \Big)} 
\Bigg)^3  \cdots\,,
\label{appc1}
\end{eqnarray}
where $u \!\cdot\! P =E_P- \vec{u}\!\cdot \! \vec{P}$.
Two quantities that enter $\delta(\cdots)$ functions in Eq. (\ref{wf13}) 
are more simply expressed in terms of the $\vec{u}$ variable, 
using Eqs. (\ref{wf17}, \ref{wf18}):
\begin{eqnarray}
\vec{p}-\vec{P}&= & \vec{u} \;\; 
\frac{ 4\,e_k^2-M^2 }{u \!\cdot\! P
+\sqrt{ (u \!\cdot\! P)^2 +(1-\vec{u}\,^2)\,(4\,e_k^2-M^2 )  } }\, , 
\nonumber \\
e-E_P&=&\frac{4\,e_k^2-M^2 }{u \!\cdot\! P 
+\sqrt{ (u \!\cdot\! P)^2 +(1-\vec{u}\,^2)\,(4\,e_k^2-M^2 )  }  }\, ,
\label{appc3}
\end{eqnarray}
while the  $\delta(\cdots)$ functions read:
\begin{eqnarray}
&&\delta\Big((p'-P)^2\Big)=\delta(1-\vec{u}\,'^2)\;
\Bigg(\frac{u' \!\cdot\! P 
+\sqrt{ (u' \!\cdot\! P)^2 +(1-\vec{u}\,'^2)\,(4\,e_{k'}^2-M^2 )  } 
}{4\,e_{k'}^2-M^2}\Bigg)^2
\nonumber \\ 
&&\delta\Big((p-P) \!\cdot\! (p'-P)\Big) = \delta(1-\vec{u} \!\cdot\!\vec{u}\,')
\;\; \frac{
u \!\cdot\! P +\sqrt{ (u \!\cdot\! P)^2 +(1-\vec{u}\,^2)\,(4\,e_k^2-M^2 ) }
}{4\,e_k^2-M^2}\;
\nonumber \\ &&  \hspace{4cm} \times \;\frac{
u' \!\cdot\! P +\sqrt{ (u' \!\cdot\! P)^2+(1-\vec{u}\,'^2)\,(4\,e_{k'}^2-M^2) }
 }{4\,e_{k'}^2-M^2} \, .
\label{appc4}
\end{eqnarray}
Some simplification occurs when the Jacobian and the $\delta(\cdots)$ 
functions are put together:
\begin{eqnarray}
&&\int \frac{d\vec{p}\,'_1}{2\,e'_1} \; \frac{d\vec{p}\,'_2}{2\,e'_2}\;
\delta\Big((p-P) \!\cdot\! (p'-P)\Big) \;\delta\Big((p'-P)^2\Big)\cdots
\nonumber \\ 
&& \hspace{1cm}=\int d\vec{k}\,'\;  
d\vec{u}\,' \;\delta(1-\vec{u} \!\cdot\!\vec{u}\,')\;\delta(1-\vec{u}\,'^2)\;
\nonumber \\  && \hspace{2cm} \times 
\frac{u \!\cdot\! P +\sqrt{ (u \!\cdot\! P)^2 +(1-\vec{u}\,^2)\,(4\,e_k^2-M^2 ) 
}
}{ 2\,e_{k'}\;(4\,e_k^2-M^2)\;
\sqrt{ (u' \!\cdot\! P)^2 +(1-\vec{u}\,'^2)\,(4\,e_{k'}^2-M^2 ) } } \cdots
\nonumber \\
&& \hspace{1cm}=\int d\vec{k}\,'\;\frac{\pi}{ e_{k'}\;(4\,e_k^2-M^2)}  \cdots 
\label{appc5}
\end{eqnarray}
In writing the last line, we took into account the relation, $1-\vec{u}\,^2=0$,
while assuming that the factor represented by dots does not depend 
on $\hat{u}'$.

\section{Normalization} 
\label{app:d}
The present appendix is devoted to the $\delta(\cdots)$ functions appearing 
in the general expression of the normalization, Eq. (\ref{wf28}) or 
also in Eq. (\ref{appa12}). By using the 
conditions implied by the different $\delta(\cdots)$ functions, one first 
successively gets: 
\begin{eqnarray}
&& \delta\Big((p-P)^2\Big)
\; \delta\Big((p-P) \!\cdot\!(p-P')\Big)\; \delta\Big((p-P')^2\Big)
\nonumber \\
&&= \delta\Big((p-P)^2\Big)
\; \delta\Big((p-P) \!\cdot\!(P-P')\Big)\; \delta\Big((p-P')^2\Big)
\nonumber \\
&&=\delta\Big((p-P)^2\Big)\; \delta\Big((p-P) \!\cdot\!(P-P')\Big)\;
 \delta\Big((P-P')^2\Big)
\nonumber \\
&&=\delta\Big((p-P)^2\Big)\;
\Bigg(\frac{u \!\cdot\! P +\sqrt{ (u \!\cdot\! P)^2 +(1-\vec{u}\,^2)\,
(4\,e_k^2-M^2 ) } }{4\,e_k^2-M^2}\Bigg)\;
\nonumber \\ && \hspace{1cm} \times
 \delta\Big(u\!\cdot\!(P-P')\Big)\;  \delta\Big((P-P')^2\Big) \, ,
\label{appd1}
\end{eqnarray}
where the last term of the equality is obtained by rewriting the term, 
$\delta\Big((p-P) \!\cdot\!(P-P')\Big)$. 

Assuming in a next step $P^2=P'^2$, 
the two last factors can be shown to be equivalent to a 
3-dimensional-$\delta(\cdots)$ function which implies  the conservation of the 
total 3-momentum. In this order, 
it is convenient to introduce the components of $(\vec{P}-\vec{P}')$ parallel 
and perpendicular to $\vec{u}$. Some detail follows:
\begin{eqnarray}
&& \delta\Big(u\!\cdot\!(P-P')\Big)\;  \delta\Big((P-P')^2\Big)
\nonumber \\
&&= \delta\Big(E_P-E_{P'}-\vec{u}\!\cdot\!(\vec{P}-\vec{P}')\Big)\;
\delta\Big((E_P-E_{P'})^2-(\vec{P}-\vec{P}')^2 \Big)
\nonumber \\
&&=\delta\Big(E_P-E_{P'}-\vec{u}\!\cdot\!(\vec{P}-\vec{P}')\Big)\;
\delta\Big((\vec{P}-\vec{P}')_{\perp}^2 \Big)
\nonumber \\
&&= \pi\;
\frac{ \delta\Big((\vec{P}-\vec{P}')_{\parallel}\Big) }{
1-\vec{u}\!\cdot\!\vec{P}/E_P} \; \;
\delta^2\Big((\vec{P}-\vec{P}')_{\perp} \Big)
\nonumber \\  &&= 
\pi\;\frac{E_P}{ u\!\cdot\! P}\; 
 \delta\Big(\vec{P}-\vec{P}'\Big)\, .
\label{appd2}
\end{eqnarray}
%

%

\end{document}